\documentclass[12pt]{article}
\usepackage{graphicx}
\usepackage[latin1]{inputenc}
\usepackage[english]{babel}
\usepackage{amsmath}	
\usepackage{amssymb}	

\begin{document}

\date{}

\title{Comment on the sign of the Casimir force}

\author{C. P.~Bachas\\ {\small Laboratoire de Physique Th\'eorique de l'Ecole 
Normale Sup\'erieure}\footnote{Unit\'e mixte de recherche (UMR 8549)
du CNRS  et de l'ENS, associ\'ee \`a l'Universit\'e  Pierre et Marie Curie et aux 
f\'ed\'erations de recherche 
FR684  et FR2687 .}  \\ {\small 24 rue Lhomond, 75231 Paris cedex, France}}

\maketitle
 
 \abstract{\footnotesize  I show that reflection positivity implies that the force between
 any mirror pair of charge-conjugate probes of the quantum vacuum  is attractive. 
This generalizes a recent theorem  of Kenneth and Klich \cite{KK} to interacting quantum
fields,  to arbitrary semiclassical bodies,  and to quantized  probes with non-overlapping
wavefunctions.  I also prove that the torques on charge-conjugate probes
tend always  to rotate them  into a mirror-symmetric position. } 

\vskip 1cm 

 
 The Casimir force \cite{Cas}
 and related electromagnetic interactions \cite{Lifshitz,IHP}\\  are the dominant forces
between neutral non-magnetic objects 
 at scales  ranging from a few nanometers to hundreds of microns. 
 These forces play  therefore an important  role both  in mesoscopic
 gravity experiments \cite{gravity}, 
 and in  the design and  fabrication of micro- 
  and nano-electromechanical systems \cite{Sri,Chan}. 
Of particular  interest  is the possibility  of repulsive interactions,  since a frequent cause of  
 malfunction of  miniature devices  is the collapse  and 
 subsequent permanent  adhesion of mechanical elements,  
known as  `stiction' \cite{stiction,stiction1}. 
 It is therefore important to understand under what
conditions such residual electromagnetic forces may change sign.

Motivated by this question  Kenneth and Klich have
recently shown  that a mirror pair of dielectric bodies always attract
\cite{KK} .   Their proof is based on
a detailed  study of the determinant  formula that enters  in the leading-order 
semiclassical calculation of the  force.  Here I want to point out that  
 this result
follows from  a  general property of quantum field theory known as
`reflection positivity'  \cite{OS,OS1}. This extends the validity of the theorem
to interacting quantum fields, arbitrary static  vacuum probes, and (with assumptions to be
discussed)  beyond the semiclassical approximation of the  probes,  as well as in the presence of
spectator bodies or material media. 
The proof  applies equally well to electrostatic and  magnetostatic forces, as to
those arising from the quantum and thermal fluctuations of the electromagnetic field.
A variant of the argument  shows that  torques always tend to rotate
a pair of conjugate probes into a mirror-symmetric configuration.  
The arguments  presented here have been used in the past to establish the concavity
of the quark/anti-quark  potential \cite{conv,conc}. The main point of the present
letter is to explain that they apply to a much broader class of
physical situations. 

\vskip 1mm
 
 Reflection positivity is a general property of Euclidean quantum field theory which
 guarantees the existence of a positive Hilbert space and a self-adjoint,  
 non-negative Hamiltonian \cite{OS}.   It can be stated as the following correlation
 inequality: 
 \begin{equation}\label{1}
 \langle\,   f\,  \Theta( f ) \,  \rangle \ \  \geq\ \  0 \ , 
 \end{equation}
 where $\Theta$ denotes the  reflection with respect to a three-dimensional  hyperplane in ${\rm R}^4$,
 $f$ is any  functional of  the fields with arguments to the left of the hyperplane, 
 and the action of $\Theta$ on $f$ is anti-unitary [i.e. $\Theta (c f) = c^* \Theta(f)$]. 
Inequality (\ref{1}) can be established easily in the path-integral formulation of
quantum field theory.
 Consider for example 
a real scalar field  defined on a hypercubic lattice $\Lambda$ which  admits $\Theta$ as  a
reflection symmetry.  Let the functional-integral  measure  be~:
\begin{equation}
d \mu =  (\prod_x d\phi_x)\,  e^{-S_\phi} \ \ \ \   
 {\rm with}  \ \ \  S_\phi = 
  \sum_{<x,y>}  {1\over 2} (\phi_x -\phi_y)^2 \ +\  \sum_{x} V(\phi_x) \ ,  
\end{equation}
where  $V$ is  the scalar potential and  $<x,y>$ the lattice links.  
We also define $\Lambda_+$, $\Lambda_0$ and $\Lambda_-$ to be the collections
 of sites and links
 that lie to the left, on, or to the right of the mirror hyperplane.  Splitting  the integration 
measure accordingly,   $d\mu = d\mu_+ d\mu_0\,  d\mu_-$  ,  one  finds 
\begin{equation}
 \langle\,   f(\phi_{x})\,  f^*(\phi_{\Theta x})  \,\rangle  \  =\  \int d\mu_0\  \,   \Bigl\vert
 \int d\mu_+  f(\phi_{x})  \Bigr\vert^{\, 2}  \  
\end{equation}
for any functional $f$ that is
defined entirely in $\Lambda_+$. The  inequality (\ref{1}), with $\Theta(f(\phi_x)) \equiv f^*(\phi_{\Theta x})$, 
 follows now from the positivity of the measure $d\mu_0$. 
  This proof only applies,  strictly-speaking,   to lattice symmetries,  but 
it extends to all mirror reflections in the continuum limit 
where rotational invariance is restored.  

\vskip 1mm

 Reflection positivity for  gauge fields can be established in a similar way  \cite{OS1}, starting
 for instance from the Wilson action
 \begin{equation}
 S_{\rm Wilson}  \, =  \,  -{1\over 2g^2}  \sum_{ P}\,   {\rm Re\  tr} \Bigl( \prod_{<x,y>\in P}  U_{<x,y>}\Bigr)\ .
 \end{equation}
Here $P$ labels the lattice plaquettes,   the product runs over the links 
of $P$ in a path-ordered way, and  the link variables $U_{<x,y>} =
 U_{<y,x>}^{\ \dagger }$ take values
in the gauge group \cite{cre}.  We define the action of $\Theta$ as follows:
 $\Theta(f(U_{<x,y>} )) \equiv f^*(U_{<\Theta x, \Theta y>} )$,
  and note  that it leaves invariant  the Wilson action. 
Inequality (\ref{1}) follows now easily
from positivity of the Haar measure,  $\prod dU_{<x,y>}$, 
by a similar argument as above.
 The extension to fermion fields is also straightforward,
provided  one defines the action of $\Theta$ appropriately  \cite{fermions,OS1}. 
  Thus (\ref{1}) is
 a general property of relativistic quantum field theories,
and in particular of Quantum Electrodynamics and of the Standard Model.
 
 \vskip 1mm 

It is important here to emphasize that the existence of a reflection operator $\Theta$ satisfying
 (\ref{1})
   is only a consequence of unitarity, and   makes no assumptions
  about the discrete symmetries  $P$, $C$ and $T$.\footnote{The issue was
raised  by Adam Schwimmer. I am grateful to him, as well as to Henri Epstein 
and John Iliopoulos,  for helping to
clarify  it.} 
This is made clear  by the following facts:

\begin{itemize}
\item  A  bounded-from-below 
potential for charged scalar fields 
respects reflection positivity, 
even though it  can lead to  explicit or spontaneous  breaking of the charge-conjugation symmetry $C$.   
\item 
 A $\theta$-term respects reflection positivity, but
 breaks explicitly parity and time reversal.  Note that 
 $\ \theta\int \epsilon_{\mu\nu\rho\sigma} {\rm tr}  F^{\mu\nu} F^{\rho\sigma}$
is odd under $P$ or $T$, but it makes an imaginary  contribution
 to the Euclidean action.  Thus the combination of 
 reflection and complex conjugation leaves  invariant the functional-integral 
 measure,  and inequality (\ref{1}) continues,  indeed,  to hold.\footnote{Defining
  the $\theta$-term on the lattice,  so as to make
the argument rigorous,  is tricky \cite{theta}.} 
\end{itemize}

 In some of the situations discussed
below,   $\Theta$ will act in the same way as  the combination $CP$.  This is however only a coincidence,
as the above examples clearly illustrate. 
 
\vskip 1mm 

There is a second point that needs clarification:  positivity of the theory
requires, strictly-speaking,  inequality  (\ref{1})  only for  reflections in the direction of (Euclidean)  time. 
 These are indistinguishable from space reflections if we consider probes of a
 relativistic vacuum.  In the presence of a   medium, on the other hand,  
 the measure need not be invariant under an operation  $\Theta$ that reflects space.  
 Only when it is, will we be allowed to extend the validity of our theorem.

\vskip 1 mm
 Let us now concentrate on functionals $f$  which, when inserted  in the path integral, 
 describe static semiclassical probes 
 interacting with the  electromagnetic vacuum.  Examples  are  a  
 superconducting shell,   a dielectric body,  or  (a  collection of) infinitely-heavy
 electric charges. The corresponding functionals read:
 
\vskip 7mm

\begin{tabular}{|c||c|c|c|}\hline
probe &conductor &dielectric&point charges\\\hline
&&&\\
$f$ &${\prod\atop {\rm P\in {\cal W}}}\  \delta (U_P - {{1}} )
$& ${{\prod} 
\atop {\rm P_0\in {\cal W}} }\  {\rm exp}( \alpha \, {\rm Re}\, U_{P_0})$ &
$\prod_j \ (U_{j})^{q_j}$ \\
&&&\\ \hline
\end{tabular}.
  
 \vskip 7mm 
 
\noindent In these formulae $U_P$ are the plaquette variables, 
${\cal W}$ is the  world-volume of the dielectric body or the conducting shell, 
$P_0$ labels  timelike plaquettes (i.e. those with one link in the time
direction), and  the $U_j$ are Wilson lines  along the worldline of the $j$th
particle whose electric charge  is $q_j$.  Note that $U_P \simeq {\rm exp}(i F_P)$,  
where $F_P$ is the electromagnetic field strength along the plaquette. 
Thus the $\delta$-functions in the first entry  impose  the vanishing
of the electric and magnetic fields  tangent, respectively normal, 
 to the conducting shell, as  appropriate.  
  Note also that the material of the dielectric
body  has been assumed isotropic and non-dispersive, with 
relative permittivity $\epsilon$ given by  $\alpha =  {(1-\epsilon) /  2g^2}$. 
We will comment on dispersive media later on. 
Note finally that one can describe geometrically-smooth 
 probes  by using  linear combinations of plaquette field strengths in the definition of
  the above functionals.\footnote{The use of lattice variables is not essential. It shows, however,  that
the relevant inequalities  survive the (non-perturbative)  regularization of the theory.}

\vskip 1mm
For static probes the functionals $f$ are time-translation invariant.
We may thus impose  periodic boundary conditions in the time direction
 ($x^0 = x^0 +\beta$)  so that
 \begin{equation}\label{5}
\langle f_1\cdots  f_N \rangle \ =\   e^{-\beta\,  E (1, \cdots ,  N)}  
\end{equation}
where $\beta$ is the inverse temperature,  and 
$E$  is the free energy of the system 
 in the presence of the probes  $\ 1, \cdots ,  N$. The 
existence of a continuum limit for the correlator (\ref{5})
raises, in fact,  some subtle  mathematical  questions. 
Careful subtraction
procedures have been proposed for Wilson-loop operators \cite{Dot} and, at the
one loop-level,  for  perfectly conducting shells \cite{BD}.  A weak assumption, verified
explicitly in the above two examples, is that  (\ref{5})  can be made finite
by a  multiplicative renormalization of the functionals,
$f_a^{\rm ren} = {\rm exp}({-\beta E_a})\, f_a \,  , $ where 
$E_a$ is the (generally divergent) self-energy of the $a$th probe.  This will be
sufficient for our purposes here, since we will focus on the dependence of the
energy on the relative position and orientation of the probes, 
rather than on their detailed composition and  shape.

\vskip 1mm

We come now to the main point of this paper. Reflection positivity implies
that $(f_1, f_2) \equiv  <f_1 \Theta(f_2)>$  is  a non-negative inner product
on the space of functionals that are defined on one side of a reflection hyperplane.
This implies,  in turn,  the
Schwarz   inequality: 
\begin{equation}\label{6}
\vert \langle  f_1 \Theta(f_2) \rangle\vert^2  \  \leq \ \langle  f_1 \Theta(f_1) \rangle
\langle  f_2 \Theta(f_2) \rangle
\end{equation}
for any two probes 1 and 2 that live on the same side of the hyperplane.
 Consider  the special case of identical probes, with the same orientation and
 with centers of mass aligned on the O$z$ axis. Let $\Theta$ act by
 reflecting  the coordinate $z$ (see figure 1).  Combining inequality (\ref{6}) 
with equation  (\ref{5})  we find~:
\begin{equation}\label{7}
2 E_{\rm mir} (z_1+z_2)\  \geq \  E_{\rm mir}(2z_1) + E_{\rm mir}(2z_2)\ , 
\end{equation} 
 where $E_{\rm mir}(z)$ is the interaction energy of the mirror pair of conjugate probes, 
 and  $z$  is the separation of their centers. It follows from
 (\ref{7})  that $E_{\rm mir}(z)$ is  a concave function, i.e. its second derivative is  nowhere positive. 
 Thus ${dE_{\rm mir}/ dz}$ is monotone non-increasing. 
 Since, furthermore, the energy  is bounded at infinite separation
  from below,  we  conclude that
 ${dE_{\rm mir}/ dz}$ is everywhere  non-negative,  so that the force is either
 attractive or exactly zero.   This generalizes  the proof of 
 reference \cite{KK} to arbitrary static external probes, 
   and beyond the one-loop approximation.\footnote{The fact that a spherical 
   perfectly-conducting  thin shell feels an outward Casimir pressure \cite{Boyer}
    does not contradict the theorem. 
   It has indeed been recognized that  separating an elastic shell into  two rigid hemispheres
    is a mathematically-singular operation, 
   which introduces  divergent edge contributions to the energy \cite{div,BD}.  } 
   An extra immediate corollary is that the binding energy of a pair of conjugate probes can never grow
   faster than linearly with distance \cite{conv}. 
 
\begin{figure}
\hskip 2cm
\includegraphics[width=0.7\textwidth,height=8.5cm,angle=0]{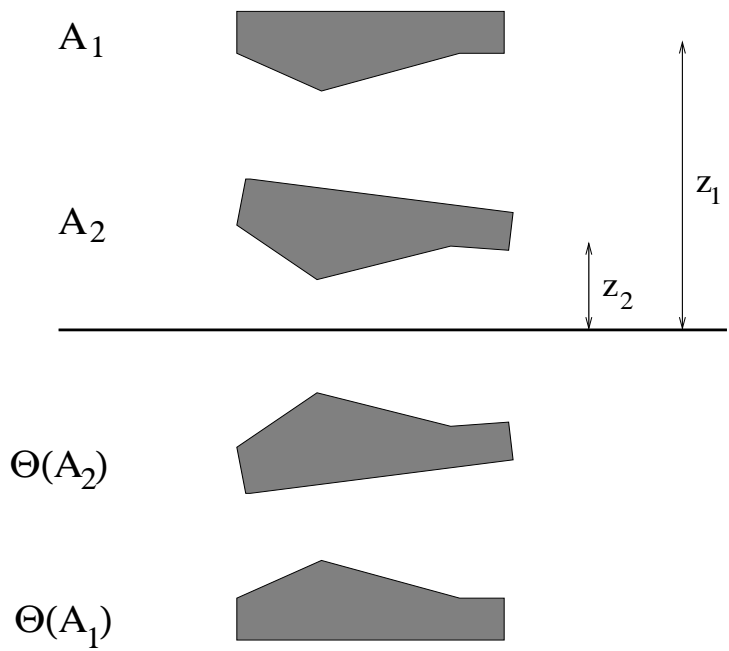}
\caption{\footnotesize Two identical probes with centers of mass
  aligned along the Oz axis, and their mirror images in the (xy)
  plane. Probe 2 is in general rotated with respect to probe 1. 
Inequalities (\ref{7}) and (\ref{10}) are obtained in the special
  limits of zero rotation, or for $z_1=z_2$.
  }
\label{f:probe}
\end{figure}

\vskip 1mm

Notice that it is crucial for the theorem's validity
 that the operation $\Theta$ involves a complex conjugation. 
  This  does not affect neutral probes, like the conducting
shell or the dielectric bodies,  
but it does flip the sign of electrostatic forces. To illustrate
 the point consider  
the classical interaction energy between a pair of electric dipoles~:
\begin{equation}\label{8}
E({\mathbf d}_1, {\mathbf d}_2 \vert  {\mathbf r}) \
 =\   {{\mathbf d}_1\cdot {\mathbf d}_2 -
  3 ({\mathbf d}_1\cdot\hat {\mathbf r})({\mathbf d}_2\cdot\hat {\mathbf r})\over \vert
  {\mathbf r}\vert^3}\ . 
\end{equation}
The action of $\Theta$  on an electric dipole is
 $\Theta({\mathbf d}) = - {\mathbf d} +  2 \hat {\mathbf n}
  ({\mathbf d}\cdot\hat {\mathbf n}) $,  where $\hat {\mathbf n}$ is a unit
 normal to the reflection plane, and the overall minus sign comes from charge conjugation. Setting 
 $\hat {\mathbf n} = \hat {\mathbf r}$ gives the interaction
 energy between a mirror pair of conjugate dipoles
\begin{equation}
E({\mathbf d}, \Theta({\mathbf d}) \vert {\mathbf r}) \ =
\   - {{\mathbf d}\cdot {\mathbf d}  + ({\mathbf d}\cdot\hat {\mathbf r})^2 \over \vert
  {\mathbf r}\vert^3}\ .
\end{equation}
The force between the pair  is thus attractive, consistently with reflection positivity. 
Note that for a magnetic dipole created by a
persistent current 
$\Theta({\mathbf m}) = {\mathbf m} -   2 \hat {\mathbf n} ({\mathbf m}\cdot\hat {\mathbf n}) $.
The theorem nevertheless continues to hold, because the interaction energy of two such 
magnetic dipoles
is minus the expression in (\ref{8}), see  for example \cite{Ja} . 
\vskip 1mm

 As a further application of inequality (\ref{6}) we can consider the dependence of the energy
 on the spatial orientation of the probes. Let 1 and 2 be  obtained from some reference 
 probe
 by rotations ${\mathcal R}_1$ and ${\mathcal R}_2$ around its center of mass
 (see figure 1).  Inequality (\ref{6}) then reads  in a self-explanatory notation
 \begin{equation}\label{10}
 2 E({\mathcal R}_1, \Theta({\mathcal R}_2))\  \geq\  E({\mathcal R}_1, \Theta({\mathcal R}_1)) +
 E({\mathcal R}_2, \Theta({\mathcal R}_2))\ . 
 \end{equation}
 It follows that $E({\mathcal R}_1, \Theta({\mathcal R}_2))$  is greater than the smaller
of the two terms on the right-hand side, so that the minimum of the energy is obtained
when the relative orientation of conjugate probes corresponds  precisely to that of 
a mirror pair.\footnote{The absolute
 preferred orientation depends on the details of  the probes,  
 and cannot    be determined by such general
arguments.} 
Since in this configuration the force is, as we saw, attractive,   
such systems will always
tend  to collapse  unless stabilized by  external torques and forces. 
 \vskip 1mm

Let us  now  turn attention to  situations 
in which a  material medium, or a spectator body,  break  Lorentz invariance even
before the insertion of external probes.  Spatial and (Euclidean-) time
reflections are no more equivalent,  
so that the theorem need not always hold.   Its (partial) validity requires that
 the functional measure be invariant under  (at least a subset of)
{\it spatial}  reflections and complex conjugation.  We have already used this fact above,
 in our discussion of   the thermal vacuum. 
The following examples further illustrate the point:

\begin{itemize}

\item  Two identical pistons
inside an infinite conducting cylinder will attract, in agreement with the conclusion
of references \cite{piston}. This follows from $\Theta$-invariance for all reflections that 
are geometric symmetries   of the cylinder. 
 
\item  The proof continues to apply in the presence of a classical homogeneous
and isotropic  dispersive  medium, as discussed already by Kenneth and Klich \cite{KK}. 
 The Euclidean Maxwell action in this case can be written as:
\begin{equation}
S_{\rm Max} =  \int d^3{\mathbf x} \int_0^{\infty} {d\omega\over 2\pi}\, \left(  \epsilon(i\omega) 
\vert {\mathbf E}(\omega, {\mathbf x})  \vert^2 
+   \vert {\mathbf B}(\omega, {\mathbf x})  \vert^2   \right) \ , 
\end{equation}
where $ \epsilon(\omega) $ is the dielectric permittivity,
which must be  analytic in the upper-half complex  plane,  and
 real and positive on the imaginary $\omega$ axis \cite{Lif}.  Although this action is
 non-local in Euclidean time, it does satisify  reflection positivity for space reflections, 
 so that  a  mirror pair of conjugate probes still  attracts.\footnote{Note that  the
 dielectric or superconducting probes discussed so far can be thought of as  special limits
 of a spatially-varying  permittivity $ \epsilon(\omega, {\bf x}) $.}

\item   A case where the theorem doesn't hold, consistently with claims
in the literature \cite{Wir},  is that of mirror probes in a Fermi sea.  Indeed a non-zero
chemical-potential term,  $\mu j^0$  where $j^0$ is the relevant  number density,  is invariant under charge
conjugation and time reflection, but not under charge conjugation and {\it space} reflection. 
This seems  to contradict our first example, since a conducting material could be after all modeled by a  
free-electron gas.  The point is, however, the following: the theorem holds to the extent that the conductor
can be described by a classical permittivity for the photon field, but doesn't apply if the electron-gas
fluctuations  become important.\footnote{ Other examples that violate the assumptions
of our theorem are periodic boundary conditions
for fermions \cite{fermions}, or the situations discussed in reference \cite{Fulling}. 
}

\end{itemize}

\vskip 1mm

This last point can be made more clear  by considering
 the extension of the theorem to quantum probes. A simple example 
is  a  charged  non-relativistic quantum particle trapped inside a potential well.
The corresponding functional of the electromagnetic field
(that replaces the Wilson line of a classical charge)  reads~:
\begin{equation}\label{12}
f(A^\mu) =  \ \int [D{\mathbf x}(t)]\,  e^{-I} \, 
e^{iq\int (A^0  dt - {\mathbf A}\cdot {d{\mathbf x}})}\ . 
\end{equation}
 Here  $[D{\mathbf x}(t)]$  is the integration measure over particle trajectories, 
\begin{equation}
I \, = \, \int_0^\beta dt  \, \left[ {m\over 2}( {d{\mathbf  x}\over dt})^2 +  V({\mathbf x}) \right] 
\end{equation}
 is the Euclidean  non-relativistic  particle action,
 and $V({\mathbf x})$ the localizing  potential.  Plugging the
 functionals (\ref{12})  in equation (\ref{5})
 gives the free energy of the combined system of particles and fields
at inverse temperature $\beta$. If we could use inequality (\ref{6}),  we would then conclude
as before  that  the force between a pair of  conjugate probes  is  attractive. 
The problem  with this argument is that the quantized particle does not, strictly-speaking,  live
 in one side of the reflecting plane.  Forcing it to do so (e.g. by adding to $V$ an artificial 
 hard-wall term) will,  however, introduce only a tiny error if the particle-mirror separation is
 much bigger than the localization length.  For a harmonic-well potential the error will, for
 instance, decay exponentially fast.   For sufficiently-large probe separations  the conclusions of the
  theorem  should thus  still apply. 
  
   We can easily extend the above argument to  collections of particles, such as the protons, neutrons
   and electrons of a heavy atom. Note, however, that because $\Theta$ acts as charge conjugation, 
    the mirror partner must be made of antimatter!  Thus this last extension
  of the theorem to truly microscopic quantum probes
 is of little relevance for  applications to realistic electromechanical  systems.

   \vskip 5mm  
   
{\small \bf Aknowledgements}:  
I thank Bertrand Duplantier and Giuseppe Policastro for useful
conversations. I also thank Israel Klich,  Erhard Seiler, Adam Schwimmer  and one of the PRL referees
for comments that  significantly improved the original manuscript.
 This research was  partially supported
 by the European Superstring
Theory  Network MRTN-CT-2004-512194.


\end{document}